# A Replica for our Democracies? On Using Digital Twins to Enhance Deliberative Democracy

Claudio Novelli[1], Javier Argota Sánchez-Vaquerizo[2], Dirk Helbing[2,3], Antonino Rotolo[4], Luciano Floridi[1,4]

[1] Digital Ethics Center, Yale University, 85 Trumbull Street, New Haven, United States
[2] ETH Zürich, Computational Social Science, Stampfenbachstrasse 48, 8092 Zürich, Switzerland
[3] Complexity Science Hub Vienna, Metternichgasse 8, 1030 Vienna, Austria
[4] University of Bologna, Alma AI and Department of Legal Studies, Via Zamboni 22, Bologna, Italy

**Abstract.** Deliberative democracy depends on carefully designed institutional frameworks — such as participant selection, facilitation methods, and decision-making mechanisms — that shape how deliberation performs. However, identifying optimal institutional designs for specific contexts remains challenging when relying solely on real-world observations or laboratory experiments: they can be expensive, ethically and methodologically tricky, or too limited in scale to give us clear answers. Computational experiments offer a complementary approach, enabling researchers to conduct large-scale investigations while systematically analyzing complex dynamics, emergent and unexpected collective behavior, and risks or opportunities associated with novel democratic designs. Therefore, this paper explores Digital Twin (DT) technology as a computational testing ground for deliberative systems (with potential applicability to broader institutional analysis). By constructing dynamic models that simulate real-world deliberation, DTs allow researchers and policymakers to rigorously test "what-if" scenarios across diverse institutional configurations in a controlled virtual environment. This approach facilitates evidence-based assessment of novel designs using synthetically generated data, bypassing the constraints of real-world or lab-based experimentation, and without societal disruption. The paper also discusses the limitations of this new methodological approach and suggests where future research should focus.

## 1. Introduction

Deliberative democracies are built upon reasoned debate and mutual justification, highlighting the importance of argumentation and dialogue in public decision-making. While there is broad consensus on the necessity of justifying laws and fostering respectful civic discourse, the effectiveness of deliberative democracy as a viable alternative to traditional aggregative approaches largely depends on careful institutional design. This design includes structural and procedural elements such as participant recruitment (who should take part), training and information provided to participants, debate structure, timing (frequency of deliberative rounds), decision-making, and monitoring mechanisms (Fung 2007).

However, selecting the most suitable institutional design to fulfill specific deliberative goals is challenging. Testing and refining such institutional designs—and assessing their impact on the quality of deliberation—poses significant challenges in both real-world settings and lab experiments, mainly due to resource constraints and limitations to generalizability and scalability.

To address these limitations, this paper explores how Digital Twin (DT) technology, a dynamic computational modeling framework, can serve as a regulatory sandbox for deliberative democracy. DTs mirror deliberative communities by integrating real-world data (e.g., demographic, behavioral, or social data) and advanced computational techniques (e.g., agent-based modeling, machine learning, or network analysis). Institutional designers, including scientists, policymakers, and public or private implementers, can use this experimental tool to simulate deliberative processes all the way from individual reasoning to collective decision-making, to test different models, generate (limited) predictive insights, explore alternatives, and refine deliberation procedures in controlled virtual environments. This kind of iterative testing is often impractical in real-world settings.

Our paper is structured as follows. Section 2 outlines the foundational principles of deliberative democracy, contrasting it with aggregative models and emphasizing its focus on reasoned discourse, mutual justification, and community-oriented engagement. Section 3 categorizes the procedural rules of deliberation and explores challenges in modeling and testing these rules using real-world and lab-controlled methods (Subsection 3.1). Section 4 introduces Digital Twin (DT) technology's core features and potential for replicating complex social systems. Section 5 examines its application to deliberative democracy as a sandbox for testing procedural rules. Subsection 5.1 expands on the needs that justify the value of such DTs. Subsection 5.2 outlines the development of a deliberative community DT, using mini-publics as a case study. Subsection 5.3 demonstrates how the DT can be used to test and refine the institutional design. Section 6 concludes the article by discussing the limitations of using DTs in this area and highlights key considerations for future research.

## 2. Deliberative democracy: the cogency of the best argument

Traditionally, the shared core of what is called deliberative democracy is the belief that the legitimate source of law derives not merely from a straightforward aggregation of people’s preferences in the face of conflicting interests, but rather from the exercise of reason and mutual respect directed toward the common good (Mansbridge 2007; Gutmann and Thompson 2009). Consequently, a key aspect of deliberative democracy involves viewing civic engagement not as the sum of isolated individuals’ will, but as a communal commitment to mutual justification and openness to persuasion.

As is well known, deliberative democracy differs from traditional aggregative democracy in various ways. From a foundational perspective, it does not rely on any single, fixed principle to determine whether a procedure or law is justified; instead, its principles are inherently dynamic and subject to ongoing revision in light of new moral and political arguments (Gutmann and Thompson 2000).

From a procedural perspective, deliberation represents a distinct decision-making paradigm. Unlike aggregative democracy, which reduces decision-making to the mere mechanistic expression of preferences through voting, deliberative democracy emphasizes argumentative complexity and aims to achieve higher standards of rationality—both substantively and procedurally (Dryzek 2002).

Scholars debate deliberation's ideal outcomes: Habermas seeks rational consensus (Habermas 1998), Elster emphasizes preference transformation through reasoned argument (Elster 1998), and Cohen supports rationally motivated consensus but acknowledges the need for voting when consensus is unattainable (Bächtiger et al. 2018, 7). Participatory Budgeting illustrates this by enabling citizens to collectively deliberate on budget allocations while relying on aggregative voting mechanisms at key stages (preference aggregation) (Yang, Hausladen, et al. 2024). Deliberation thus serves multiple goals, including epistemic improvements in collective knowledge (Estlund 1993; Nino 1996; Marti and Besson 2006) — all of which share the idea that outcomes should be the most rationally justified under conditions of disagreement.

Where is deliberative democracy practiced? Formal state institutions—parliaments, executives, and courts—contain deliberative elements in the form of debate and discussion. However, they are not the purest exemplars of deliberative democracy, being more marked by strategic communication and the pursuit of individual or group interests than by genuine reason-based mediation. In such settings, the best-positioned argument still prevails over the most rationally robust one (Cohen 2021). Courts may be somewhat more deliberative among these institutions, but the judicial recruitment process is often not democratic.

More explicitly inspired by deliberative standards are non-partisan forums—often called mini-publics when they involve relatively few participants, and citizens' assemblies or deliberative polls when larger numbers are involved. In both formats, participants meet face-to-face and typically develop a recommendation or report on a policy issue (André Bächtiger et al. 2018; Curato and Böker 2016; Fournier 2011). Topics these bodies address include the risks and promises of new technologies, climate change, electoral system reform (e.g., the British Columbia Citizens' Assembly), urban development, and fiscal policies (Ganuza and Francés 2012; Fishkin 2013; Landemore 2021; Willis, Curato, and Smith 2022).

## 3. The Institutional Design of Deliberative Democracy: A Three-Rule Framework

Deliberation differs from free discussion, as its success heavily depends on institutional design (Grönlund and Herne 2022, 170). This design is shaped by a series of rules that govern different stages of the process: before, during, and after the deliberation. These rules determine critical aspects, such as who participates, how the process unfolds, and what is done with the outcomes (Parkinson and Mansbridge 2012; Fishkin 2009; Gutmann and Thompson 2000). Together, they create the framework within which deliberative democracy operates. These rules can be classified as follows:

a) *Pre-deliberation rules* concern how to set up the deliberation process. These rules define the composition of participants —ranging from random selection (Gastil and Richards 2013) or targeted stakeholder inclusion to purely voluntary participation. They also address agenda setting, whether the goals are predetermined, defined by facilitators, or collaboratively decided by the group. Equally important is the informational preparation of participants, who may be equipped with expert presentations, background reading materials on the policy problem (Weeks 2000), or access to open data to ensure they are well informed about the topic under discussion (Ruijer et al. 2024). This preparatory phase lays the foundation for meaningful and fair deliberation.

b) *Discussion rules* come into play once deliberation begins, governing how the process is conducted and how participants engage with one another. For example, the deliberation format may comprise a single session, multiple rounds, or iterative discussions that build on previous exchanges. Speaking turns can be carefully structured—whether controlled by a facilitator, constrained by time limits in open-floor discussions, or managed through queued requests. Similarly, decision-making approaches vary, ranging from consensus-building and majority voting (Cohen 2021) to methods like deliberative polling, which emphasize informed opinion (Fishkin and Luskin 2005). The role of facilitators —crucial for guiding the deliberation—can differ depending on the chosen approach, whether they aim for neutral mediation, prioritize specific agenda items, or rotate facilitation duties among participants (Escobar 2019; Moore 2012). Other aspects, such as how arguments and rebuttals are handled – e.g., open or paired rebuttals—mechanisms for resolving conflicts—e.g., structured dialogue or majority rule in stalemates—and criteria for determining "winning" arguments—e.g., iterative refinement or argument scoring/ranking—are all carefully designed to ensure a productive process (André Bächtiger and Parkinson 2019; Moshman 2020).

c) *Post-deliberation rules* address what happens after the deliberation concludes, focusing on outcomes and their follow-up (Gutmann and Thompson 2000). These include documenting results, such as summary reports or actionable recommendations, and gathering participant feedback through surveys or anonymous evaluations (Hartz-Karp 2005). Depending on the objectives and privacy considerations, the resulting reports may be made public or kept confidential. Monitoring the implementation of decisions is another crucial element, often achieved through public updates, audits, or similar accountability measures. These steps ensure that deliberation does not end as a theoretical exercise but has a tangible impact on policy or practice (Stark, Thompson, and Marston 2021).

The content of these rules and the choice between different options are heavily influenced by two key assumptions: (i) the model of deliberative democracy being considered—whether it leans more toward liberal or communitarian principles (Forst 2001)—and (ii) the approach used to assess the quality of deliberation itself. For instance, a more communitarian model might prioritize consensus-building and mutual understanding, whereas a liberal model might emphasize fair representation and structured debate. Regarding the quality of deliberation, various metrics have been proposed, all broadly inspired by core values historically associated with deliberative democracy, such as inclusiveness, respect, equality, and reason-giving, though each metric emphasizes these values differently. Some of the most widely recognized metrics include discourse quality indices (Steenbergen et al. 2003; Bächtiger, Gerber, and Fournier-Tombs 2022), the group deliberative reason index (Niemeyer and Veri 2022), levels of justification (Bächtiger et al. 2010), the listening quality index (Scudder 2022), and the extent of respectful engagement (Mansbridge et al. 2012).

A full exploration of these evaluation metrics and their ideal applications lies beyond the scope of this paper. Instead, we address a cross-cutting issue: testing the feasibility of different deliberative processes and examining their impacts. The outcomes of such feasibility tests should inform which procedural rules are most suitable among the alternatives above. For instance, we might ask whether a particular setting calls for placing a higher priority on providing information or strengthening the moderator’s role. This line of inquiry applies to various expected outcomes of deliberation and the different metrics chosen to evaluate them.

### 3.1. The challenges of modeling and testing deliberative democracy

Because deliberative democracy is primarily a normative theory and lacks a single universal model for deliberation, determining the most suitable methods to model and test its pre-deliberation, discussion, and post-deliberation rules is especially challenging. Nevertheless, empirical approaches remain relevant (Grönlund and Herne 2022, 166).

One way to study deliberation is to monitor real-world settings, where an actual community deliberates on a concrete issue. Alternatively, researchers can employ controlled experimental designs, such as lab-in-the-field experiments, in which a sample of the population interacts (online or in person) in a staged deliberative format, or scenario experiments (Werner and Muradova 2022). These methods enable researchers to study causal mechanisms—such as how participation in a mini-public influences individual opinions (Setälä, Grönlund, and Herne 2010) or how different voting methods affect the perceived legitimacy of collective decision-making (Hausladen et al. 2024) — by isolating and testing specific variables, such as discussion format and participant composition while holding other factors constant. This approach helps determine the causal impact of deliberation (Grönlund and Herne 2022; Kingzette and Neblo 2022).[1]

A key advantage of these experimental methods is that they allow researchers to focus on individual components of the deliberation process—and their governing rules—to measure their effects more precisely. This, in turn, enhances the replicability of research across different contexts, including various countries. Nevertheless, both real-world and lab-in-field experiments have their limitations:

1) *Replicability*. Real-world deliberative settings often involve unique socio-political, cultural, and institutional features that make them difficult to reproduce and impede the testing of causal mechanisms across multiple contexts. In addition, real-world deliberation depends heavily on political will, funding, or community enthusiasm, which may not be consistently available for subsequent iterations. In contrast, lab-in-the-field experiments offer greater replicability by allowing researchers to follow standardized procedures across different samples (Grönlund and Herne 2022). However, this advantage comes with a trade-off: controlled experiments often simplify or strip away the complexity of real-world contexts to ensure replicability. While this enables researchers to isolate and test specific variables, it raises questions about the external validity of the findings—that is, the extent to which the results can be generalized to real-world settings.

2) *Generalizability*. Both real-world deliberations and controlled experiments encounter significant challenges in generalizing their results—that is, in applying insights from specific cases or studies to broader, more diverse, or real-world contexts. Real-world deliberations are typically unique, context-specific events, making it challenging to transfer lessons across different policy domains (Levine 2005; Parkinson 2006)—for instance, an environmental policy issue in one region may not translate well to a healthcare policy in another. Likewise, experimental designs often rely on small, carefully selected samples (e.g., mini-publics) that, while representative in some respects, may not capture the full diversity of perspectives, interests, and backgrounds found in larger populations. This limitation also underscores scalability challenges, as expanding the number of participants or iterations is often impractical within these methodological frameworks (Friedman 2006). Finally, experimental studies tend to focus on short-term outcomes (such as immediate opinion changes), whereas deliberative democracy seeks long-term gains in civic engagement, trust, and legitimacy.

[1] A 2012 Finnish study on enclave deliberation and group polarization tested how like-minded versus mixed-opinion discussion groups affected opinion shifts on immigration policy (Grönlund, Herne, and Setälä 2015).

3) *Flexibility to iterations and follow-ups*. Identifying the "best fit" for a deliberation process often involves multiple testing cycles, learning from feedback, and refining the approach iteratively. A key challenge in real-world testing is that outcomes are only observable after the process is complete, leaving no opportunity to refine the procedure and assess how those refinements might improve the deliberation. In other words, once a real-world deliberation begins, it is often difficult to pause, modify, and restart the process.

   In theory, this issue may be less pronounced in laboratory experiments. However, researchers often maintain fixed procedures to ensure the reliability and consistency of their results. This rigidity can limit opportunities for iterative improvements or real-time adaptations, even if certain aspects of the deliberation design prove ineffective (Lee et al. 2022). Furthermore, if participants become aware of changes or adaptations during testing, their behavior may be influenced—a concern that leads directly to the next problem.

4) *Observer Effect*. A specific challenge in empirical experiments, particularly in lab-controlled settings compared to real-world scenarios, is the so-called observer effect or Hawthorne effect (Oswald, Sherratt, and Smith 2014). This phenomenon refers to the possibility that participants in a deliberative democracy study may alter their behavior when they know peers, media, or researchers are observing them (Gastil 2000; Carpini, Cook, and Jacobs 2004). For example, individuals may feel less inclined to express dissent or voice unpopular opinions under observation and more inclined to agree with dominant opinions or authority. In lab or online experiments, participants are typically more aware that they are part of a study, which can compromise the authenticity of the deliberation process. This heightened awareness may lead them to behave differently than in real-world political or civic contexts, where the pressures and dynamics of observation are often less explicit. As a result, observation may also change the outcomes of deliberation, e.g., inducing conservative decision-making.

5) *Participant compliance, retention, and engagement*. Ensuring that participants (or facilitators) adhere to agreed-upon protocols and remain engaged throughout the process is a significant challenge in real-world deliberation and experimental settings. Even with clear guidelines and skilled facilitation, logistical disruptions or a lack of perceived benefits can lead to protocol deviations or high dropout rates, jeopardizing data integrity and representativeness.

   Time constraints often exacerbate this challenge. While it is generally easier to enforce protocols (e.g., turn-taking rules, discussion time limits) and maintain engagement in lab or online settings, not all participants may fully comply, particularly if the task is lengthy or perceived as low-stakes. Additionally, since participants often receive only a one-time incentive, dropout rates can be high after initial recruitment, thereby endangering sample sizes and the representativeness of the study.

6) *Resource and time constraints*. Conducting repeated or large-scale experiments to refine procedural rules is resource-intensive, requiring substantial financial and human investment (Iyengar et al. 2003). Effective testing involves organizing deliberative events, recruiting and compensating participants (if applicable), training facilitators, securing venues, providing expert materials, collecting and analyzing data, and managing logistics. Deliberative processes often span multiple sessions and require follow-ups, which can be challenging, given the limited availability of participants and organizers. Although experiments are generally less costly than large-scale real-world deliberations, they still demand funding for recruitment, platform development, and participant incentives, particularly when conducted across diverse contexts.

The following sections examine how Digital Twin (DT) technology can address these limitations.

## 4. The Digital Twin (DT) Technology

Institutional change should ideally be based on evidence such as empirical, measurement, or experimental data. In the following, we propose a combined approach. Using available data, we suggest running many what-if scenarios to identify promising democratic frameworks, which would then be tested in experimental settings before upgrading real-world political systems accordingly (Aguilar et al. 2024). Such an approach makes an experimental approach scalable. Being partly simulation-based and partly experimental, our approach can also fix some of the shortcomings of solely computational methods, such as recently proposed generative AI applications to democracies. Identifying the most promising democratic frameworks before experimenting with and implementing them can undoubtedly benefit from computational experiments. For example, one could consider improving voting rules or changing deliberation procedures (Yang and Bachmann 2025).

One may use suitable digital twins (DTs) for the computational experiments proposed above. A DT is a dynamic, computer-based model replicating a physical entity—such as an object, process, person, or human interactions—using real-time data to mirror its behavior, performance, and evolution (Barricelli, Casiraghi, and Fogli 2019). The concept was initially coined for life cycle management in manufacturing and aerospace (Grieves 2015), echoing early visions of digital worlds. Over time, DTs have gained interest in more complex settings that are not as easily predictable: e.g., manufacturing, healthcare, and smart cities.

DTs stand apart from static models by continually integrating data from their physical counterparts and surrounding environments through IoT, sensors, AI, and predictive analytics (Fuller et al. 2020). Constantly synchronized with its physical twin through bidirectional data flows and feedback loops, a DT monitors ongoing processes and anticipates future trends, including potential damages and failures. This continual update cycle—often referred to as the 'twinning rate'—involves measuring the real-time state of the physical entity and replicating those parameters in the virtual environment. Vice versa, enabling the virtual environment to inform and change the physical environment, so that both states remain as close as possible (Jones et al. 2020, 42–43). Ultimately, DTs enable scenario testing, inform decision-making, and support proactive interventions to enhance the real-world system they represent (Barricelli, Casiraghi, and Fogli 2019, 167656). Suitable DTs may also be used to study complex dynamics, the emergence of collective behavior, (systemic) risks, and opportunities of novel democratic designs (Caldarelli et al. 2023).

The success of DTs relies on serving a purpose, being trustworthy, and functioning effectively. Therefore, the first step is to define the purpose of the DT, which can range from real-time monitoring and predictive maintenance to more exploratory what-if analyses. Next, robust data infrastructure planning and collection are essential for maintaining reliable, real-time information flow. In this phase, practitioners identify and gather data sources—such as historical data, real-time sensor readings, or external datasets—depending on the physical or social entity (Jones et al. 2020). Sensor data might include temperature, pressure, location, or speed measurements for physical objects such as vehicles.

By contrast, modeling a complex social system—such as a deliberative democracy—may also be based on data drawn from social media, online forums, administrative records (e.g., demographic data), and qualitative surveys or interviews (Franco-Guillén, Laile, and Parkinson 2022). Because social data is typically heterogeneous and may include text, images, and geospatial information, it often requires preprocessing with techniques like natural language processing (for text) or image recognition (for visuals). This complexity reflects humans' inherently multi-layered nature and interactions, creating a multi-level complex system (Helbing and Argota Sánchez-Vaquerizo 2023, 83). After cleansing and normalizing these varied inputs, the data undergoes Extract, Load, Transform (ELT) processes to prepare it for storage and analysis. This final step ensures that the DT's software can interpret and use the data effectively, continuously refining the model and delivering actionable insights into the real-world system it mirrors through interface, human-mediated action, or even actuators (Jones et al. 2020).

Once the primary objective is defined and the relevant data is collected, the next step is to develop the core model. The specific configuration of this model depends on the nature of the replicated entity, including its components, interactions, and constraints. This explains the wide range of modeling techniques enabling DTs. Geometric modeling (e.g., CAD, point clouds) is key for representation and spatially supporting interaction layers. Behavior and dynamics can be modeled in a continuum of approaches from purely physics-based to purely data-driven. Physics-based modeling derives from first principles and classical mechanics, typically governed by equations and relying on high-fidelity methods (e.g., Finite Element Modeling—FEM—, Computational Flow Dynamics—CFM—) (Hinchy et al. 2020; Thelen et al. 2022). Equally important are the social and behavioral sciences, as human interactions and decision-making processes can critically shape system outcomes. Computational social science brings these different modeling approaches and disciplines together, bridging social processes with both physics-based and data-driven frameworks. Finally, complexity science provides theoretical tools for capturing emergent phenomena, feedback loops, and multi-scale interactions.

If we aim to twin multi-instance settings (e.g., agents, states, interactions, levels, scales), the modeling approach might focus on relations and interactions such as Agent-Based Models (ABMs), Graph-based Models, Discrete Event Simulations (DES), System Dynamics (SD), or their combination (Orozco-Romero, Arias-Portela, and Saucedo 2020; H. Qiu et al. 2023). In an ABM, autonomous "agents" are defined with particular behaviors, rules, or preferences, and their micro-level interactions can lead to emergent, bottom-up, system-wide phenomena—such as shifts in collective opinion or the formation of new coalitions (Bonabeau 2002). Graph-based modeling, by contrast, centers on the ties among instances, making it possible to encode domain-specific knowledge, suitable for hybrid modeling techniques (knowledge graphs, Bayesian networks, Graph neural networks), that also make semantic representations possible ( Zheng, Lu, and Kiritsis 2022; Listl et al. 2024). Also, graph representation may allow one to identify key influencers, detect subgroups, or reveal communication patterns in social structures through Social Network Analysis (SNA) (Borgatti et al. 2009) or social influence models for opinion formation (Chacoma and Zanette 2015). DES simulates discrete events over time—often under assumptions like rational choice or bounded rationality—to examine how incremental changes (e.g., a new message or vote) can influence the system's overall trajectory (Goldsman and Goldsman 2015). Finally, SD can handle stocks, flows, time delays, and feedback loops at an aggregate level to analyze how collective behaviors or broader interventions, such as policy changes, may affect the entire system.

At a higher level, from a systems modeling perspective, developers can adopt a multi-model architecture or "co-simulation" framework to integrate multiple modeling approaches (Borshchev and Filippov 2004) within a single DT environment for different instances.

Once the core model is built, its performance is typically refined through an iterative calibration loop in its life cycle: outputs are compared to real-world results, also with the help of metrics evaluating the quality of deliberation (see Section 3), and parameters are adjusted based on discrepancies. In many cases, developers will add predictive modeling or social simulation tools (e.g., discrete event simulation) to forecast future states, test various "what-if" scenarios, anticipate outcomes, and inform decision-making (Agalianos et al. 2020). These models may incorporate real-life data—such as traffic flows—to calibrate and adjust simulated behaviors in real-time, as part of the twinning process, where sensor fusion, adaptive, probabilistic, and uncertainty modeling (e.g., in information exchange) play a crucial role (Thelen et al. 2022). The overarching aim is to transform incoming data (e.g., from sensors, social media, and databases) into actionable outputs like predictions, insights, or recommended actions for decision-making or to (semi-)automate changes in the mirrored physical environment. The final step involves completing the user interface—enabling users to issue commands or make adjustments in the digital environment—including mixed-reality settings (Argota Sánchez-Vaquerizo 2024) and selecting a suitable DT platform for deployment.

Throughout the DT development process, AI can play numerous roles, from data processing and model-building—e.g., using Machine Learning to map unknown relations between inputs and outputs from historical data—to predictive tasks—e.g., time-series analysis—and scenario testing or optimization—e.g., sensitivity analysis and reinforcement learning. AI also supports continuous monitoring of system states, further enhancing the adaptability and utility of the digital twin (Rathore et al. 2021). More recently, new foundation models that are trained by using Deep Learning (DL) based on vast amounts of data open up new possibilities for analysis, modeling, and beyond. Their great performance in generative tasks can ease the retrieval of data, their interpretation, modeling, and interfacing with digital twins.

## 5. Testing Deliberative Democracy Through Digital Twins

Having briefly introduced DT technology, we now examine its potential application to deliberative democracy. Specifically, we aim to determine whether DTs can more effectively test the procedural rules introduced in Section 3 than conventional real-life or lab-based experiments. We propose to develop an experimental approach that uses a DT of a hypothetical or actual deliberative community, whether currently active or historical, as a regulatory sandbox to test procedural rules and identify those best suited to achieve the desired outcomes. In what follows, we first clarify what is the point of using a DT in this context (Section 5.1), then we outline how one might develop a DT of a deliberative community (Section 5.2), and finally discuss how it could be employed (Section 5.3).

### 5.1. The Case for Digital Twins in Deliberation Research

Before exploring in detail how a DT of a deliberative community can be developed and deployed, it is essential first to clarify precisely why such technology is useful and what specific purposes it serves in deliberative democracy. We distinguish three interconnected levels of justification.

At the first, practical level, the DT's purpose is not to replace real-world deliberation, but to supplement it by systematically testing a variety of what-if scenarios with different procedural rules and configurations without real-world risk. Yet, given deliberative democracy's fundamentally normative nature, one must ensure that such procedural 'enhancements' are ethically and practically suited.

This brings us to the second level: the normative justification. A DT can indeed be aligned with deliberative values in both design and implementation. In fact, during the design phase, the construction of a DT can itself embody a deliberative process. Stakeholders—including citizens in contexts such as urban planning—can collaboratively define the objectives of the digital twin, along with relevant evaluation metrics or benchmarks. Accordingly, rather than creating a technocratic "black box" that obscures interactions between regulators and citizens, the DT should serve as an open, participatory, regulatory sandbox. Citizens then gain opportunities to become active co-designers of laws and policies rather than being passive recipients.

During the implementation phase, DT simulations enable evaluations of whether particular procedural rules effectively achieve deliberative democratic goals such as diversity, inclusion, active participation, trust, and legitimacy.

Methodologically, while existing empirical studies provide fundamental insights (Rowe and Frewer 2000; Smith 2009) (REFS), DT can precisely quantify the impact of these principles under varied conditions.

DT simulations add value beyond what empirical studies alone can achieve. While empirical studies yield invaluable insights, they are inherently limited by their context-specificity, logistical complexity, ethical constraints, and difficulty in isolating interactions among multiple variables simultaneously. Even minor variations in conditions or parameters might significantly impact outcomes and cause undesirable disruptions. Moreover, the generalizability of empirical results may be limited (as mentioned in Section 3.1).

To better illustrate this point, let's assume that we have some empirical evidence, e.g., that iterative deliberation reduces polarization, as participants are repeatedly exposed to opposing views. This evidence is, however, coupled with the observation that, although iterative sessions lead to deeper dialogue, this requires significantly more time and resources. DT simulations can extend this evidence and quantify how the depth of argumentation or polarization reduction evolves across each deliberative round. More importantly, DTs could reveal non-linear patterns—for example, discovering diminishing returns or even counterproductive effects (a "backlash") after a certain number of iterations. For instance, a DT may indicate that, while initial deliberation rounds significantly reduce polarization, adding further rounds beyond a specific point yields little additional benefit and might even re-polarize participants due to fatigue or entrenched resistance to repeated confrontation. Such insights could uncover subtle interactional dynamics that empirical studies might miss, whether due to temporal constraints, missing data, the high costs of replication, or the fact that certain patterns were not initially hypothesized or measured. We should stress here that while DTs can supplement empirical research by filling these gaps, they need not replace empirical observations: instead, they can serve as a preliminary exploratory tool, generating hypotheses that can later be verified empirically (Aguilar et al. 2024), or to expand and fine-tune evidence.

In short, DT simulations enable the systematic testing and refinement of empirical insights under controlled (virtual) conditions. This process optimizes deliberative procedures by highlighting subtle interactions or unforeseen dynamics that empirical methods alone might miss or find challenging to isolate. Furthermore, DTs uniquely facilitate sensitivity analyses of scenarios that would be practically difficult or ethically problematic to test in real deliberations—such as extreme polarization, group dynamics under intense stress, or rapid procedural changes impacting participant engagement and deliberative outcomes.

### 5.2. How to develop a DT of a deliberative community: the case of mini-publics

A DT of a deliberative community is a computational model designed to replicate the structure, internal dynamics, and behaviors of a deliberative community, whether hypothetical or actual. Unlike a basic simulation model, a DT is a dynamic,

"living" virtual counterpart, which may continuously be updated with real-world data. This real-time linkage to operational data enables the DT to test scenarios, potentially to predict outcomes, and to support decision-making processes (Boschert and Rosen 2016, 61).

Human DTs have been recently suggested as an algorithmic representation of individuals that can participate in political processes, as well as prescriptive agents exploring aspects of participation (García-Marzá and Calvo 2024). Creating a deliberative community requires collectivizing human digital twins, that is, replicating individuals or singular entities as statistical or behavioral archetypes and then interconnecting them into networked systems. In such collective architectures, individual DTs are not isolated but do dynamically interact, thereby mirroring social and communal behaviors.

Integrating multiple human DTs into collective structures enhances the predictive power of the deliberative community DT by capturing emergent behaviors that result specifically from group-level interactions. Moreover, collective DT networks can reveal systemic properties such as resilience, vulnerability to misinformation, or consensus dynamics, providing richer insights into how deliberative communities evolve over time and under various scenarios.

It's important to note that individual human DTs do not necessarily need to be exact digital replicas of specific persons. Depending on the intended use case and simulation goals, DTs may selectively represent only a subset of key sociodemographic attributes, behaviors, or cognitive features relevant to the scenario. For instance, if the goal is to simulate collective decision-making or opinion dynamics, DTs should primarily include attributes influencing deliberative behavior, opinion change, or social influence. At the same time, human behavior, especially in deliberative contexts, often emerges from latent psychological, emotional, or cognitive properties that may initially appear unrelated but prove critical under certain conditions. Thus, an argument favoring more comprehensive individual replication, including seemingly non-relevant attributes, is justified to achieve a highly accurate and reliable predictive model. Yet, such extensive replication significantly increases complexity, computational requirements, and the difficulty of interpretation. Modelers must therefore carefully consider a pragmatic compromise: strategically including latent features with potential behavioral significance, supported by empirical evidence or theoretical justification.

Since democracies are normative frameworks, DTs for democratic deliberative processes require a value-sensitive approach (Helbing et al. 2021). In contrast to a mass-surveillance-based approach to create an as-detailed-as possible DT including personally identifiable individuals with real characteristics, we propose to use privacy-preserving synthetic populations with statistical fidelity. This means understanding human DTs as statistical constructs rather than as direct replacements for real individuals, mitigating ethical concerns regarding as-detailed-as-possible DTs (Fontes, Carpentras, and Mahajan 2024). This approach transforms individual sociodemographic data into equivalent synthetic populations of human DTs that preserve the same statistical features, patterns, distributions, correlations, and interactions of the original population, through advanced privacy-preserving generative models (Chapuis, Taillandier, and Drogoul 2022; Hoene et al. 2025) or special data collection methods that avoid individual attribution. Such synthetic human DTs enable social DTs that are based on smart agent-based modeling, which allow for realistic generalized simulations of democratic processes without relying on specific individuals.

Once a particular conception of deliberative democracy and associated quality metrics have been defined and the DT's objectives are set, the next step is data collection. This typically includes socio-demographic information (such as age, education, ethnicity, and other socio-economic indicators) and critical behavioral and interaction data. Sources range from mainstream social media platforms (e.g., Facebook Groups, LinkedIn, X, Threads, Reddit, or even Wikipedia for tracking community interactions) to specialized civic engagement tools like Decidim, Citizen Lab, Polis, or Ethelo (Shin et al. 2024). While the former often provides more data, the latter tends to yield more structured and purpose-focused interactions. However, specialized platforms—like social media—can introduce biases by attracting users who are already civically engaged.

Additional data can be gathered from official government channels—such as parliamentary or congressional debate transcripts, voting records, press releases, and party communications. For an overview of how big data analysis might be applied to deliberative democracy experiments, see (Franco-Guillén, Laile, and Parkinson 2022).

Data for a deliberative community DT can be collected in several ways. Many civic engagement platforms provide APIs or export features for user interactions, discussion posts, and voting records. Where no official API or export option exists—often the case on forums and social media—Web crawlers or scrapers may be used to collect and periodically update posts, comments, and participation metrics (Franco-Guillén, Laile, and Parkinson 2022, 236). However, this approach may raise legal concerns, necessitating compliance with relevant laws, regulations, and the terms of service of the targeted websites. Any non-digital data must be gathered manually.

These data streams feed into the DT to keep it current. However, the frequency of updates should be balanced against computational costs, and calibration and validation—particularly challenging for social systems—must be done carefully.[2] In some instances, frequent updates may be unnecessary if the DT's primary goal is to test procedural rules under controlled conditions.

[2] Calibration and validation are distinct processes. Calibration fine-tunes a model's parameters to align its outputs with observed data from the current state of the physical system, improving its representation. Validation, however, evaluates the model's accuracy and reliability by comparing its outputs to independent datasets not used in calibration, testing its ability to predict system

A major concern in computer-assisted text analysis for the social and political sciences is ensuring the data are as clean and bias-free as possible—or at least understanding how missing or misreported data could affect the DT's predictions. Equally important is considering the context and interpretation of the data. Mitigation strategies should be documented and implemented to address these issues (Lucas et al. 2015).

In this context, Structured Topic Modeling (STM) offers a robust method for analyzing large-scale textual data. By incorporating relevant metadata—such as age, education, platform type, or time—alongside textual content, STM allows researchers to systematically identify and track discussion topics across different subgroups or contexts (Franco-Guillén, Laile, and Parkinson 2022, 238). Similarly, argument mining techniques like Argument Structure Analysis have been developed to extract debate structures in online deliberation. These techniques use natural language processing (NLP) to analyze how discussions evolve, identify controversial issues, and pinpoint conflicting viewpoints within deliberative exchanges (Lawrence et al. 2017).

These methods offer insights into which topics resonate with specific demographic segments or user communities and how discourse shifts over time. Developers can refine calibration and validation processes by integrating these findings into the DT. For instance, if real-world data show that specific topics appear more frequently than the DT predicts, the parameters can be calibrated to align more closely with STM results. The ensuing iterative feedback loop helps reduce bias, address blind spots, and ultimately produce a more accurate representation of collective deliberation.

The modeling of a DT is among the most complex stage. A DT system should integrate multiple modeling paradigms—as emphasized by the pluralistic modeling approach (Helbing 2010)—to represent deliberative scenarios better. This plurality of models also favours multi-instance architectures, where different modules or components are modeled, coordinated, and orchestrated through different simulation paradigms. These paradigms include simulations based on existing (historical) data and predictive models that assess how changes in variables or rules are expected to impact the simulation outcomes.

Among these approaches, Agent-Based Modeling (ABM) may be particularly useful. ABM has been studied in the broader democracy and decision-making literature (Qiu and Phang 2020) and has received some attention in deliberative democracy (Lustick and Miodownik 2000; Lee et al. 2022; Butler, Pigozzi, and Rouchier 2019). In ABM, each community member or facilitator is represented as an "agent" characterized by key sociodemographic attributes such as age, income, and education; these agents are defined by decision rules, preferences, behaviors,

---

behavior under varying conditions. For example, validation might assess how well the model forecasts outcomes in scenarios like changes in deliberative community composition or voting system variations.

and engagement levels (including the propensity to vote). ABM further models interactions among agents, capturing processes like opinion diffusion, coalition formation, and resource allocation. Its ability to represent heterogeneity among agents and simulate emergent phenomena makes ABM a powerful tool for studying the dynamics of deliberation. ABM capabilities are particularly boosted by embedding complex behavior into its agents to perceive, process, and react more effectively and realistically in the environment.

Recently, the development of Large Language Models (LLMs) foundation models open up new possibilities for improving knowledge retrieval, adaptability, and reasoning of agentic AI, approximating behavior expected in humans (Park et al. 2023), particularly when combined with more reliable information access (e.g., RAG) (Zhang et al. 2024). This enables the creation of human DTs that can mimic not only observable behavior, but also cognitive and emotional processes, providing a richer base for modeling deliberative settings. The ABM populated with autonomous agents that encapsulate multiple modules for mirroring different aspects of human cognition (e.g., short and long-term memory, social and action awareness, talking, etc.) are not only able to exhibit plausible human behavior, but the resulting virtual social system may also show emergent features and evolution that may be related to human societies (Altera et al. 2024), including more effective task solving through collaboration (Qian et al. 2025). Distinguishing between a reliable simulation model and a realistically looking deep fake, however, remains a challenge.

A different approach to modeling deliberative processes focuses on aggregate relationships and dynamic interactions. System Dynamics (SD) is well-suited for exploring macro-level trends, as it models complex systems through feedback loops and time-dependent behaviors (Bala, Arshad, and Noh 2017).[3] In contrast, Social Network Analysis (SNA) maps social relationships and communication flows, making it particularly useful in deliberative democracy contexts for identifying influential communities that shape debates (da Silva, Ribeiro, and Higgins 2022).

Complementary to traditional social dynamics studying opinion formation, recent developments with LLMs have also been used to study the emergent properties of their resulting social network regarding information propagation, opinion dynamics, and echo chambers (Zheng and Tang 2025). LLMs have facilitated exploring how processes, such as voting rules, can be improved for participatory budgeting. Some of these approaches allow for multiple, complementary solutions rather than a one-size-fits-all solution, which can increase satisfaction among participants. It may also be beneficial to replace classical deliberation formats with new forms of co-creation, such as "re-mixing"

[3] However, SD typically abstracts away micro-level details—individual agents, local network structures, etc.—so when applying it to deliberation or opinion dynamics, modelers must be comfortable with the loss of fine-grained resolution (Bala, Arshad, and Noh 2017).

(Carpentras, Hänggli Fricker, and Helbing 2024). Furthermore, LLMs have been proposed to simulate the decision-making and interactions of diverse personas (Yang, Dailisan, et al. 2024). LLMs can also be used as moderators. Other generative AI solutions may support the co-creation of integrated ideas and solutions or be applied to illustrate the consequences of certain decisions and ideas.

Another technique, Discrete-Event Simulation (DES), simulates distinct events within a deliberation process—such as argumentative rebuttals, the emergence of consensus, or voting phases—to assess their impact on community outcomes (Charalabidis 2011). While DES is unconventional in this field (typically used in logistics and service operations), its application to deliberative processes is theoretically feasible and could provide novel insights.

Regardless of the chosen modeling approach—or a combination of approaches—the model must be dynamic, being updated in response to agent interactions and external influences (e.g., using RAG). For example, it should predict how often and to what degree agents change their opinions based on peer interactions, exposure to news, or official statements. Machine Learning (ML) or other AI techniques can be integrated to forecast sentiment or voter turnout changes based on historical patterns to enhance predictive capabilities.[4] These predictions can then be fed back into the model (e.g., an ABM or SD model) to refine its accuracy and realism.

Once developed, domain experts, community leaders, or academic researchers must calibrate and validate the model to accurately reflect real-world processes. This step is critical for establishing the model's credibility. Additionally, iterative refinement may be necessary to incorporate new data, address feedback loops, and improve the model's precision over time. The "black box" character of state-of-the-art AI based on deep learning (DL) challenges efforts for explainability and accountability. However, combining these models with cognitive architecture designed to align with human cognitive processes may help to mitigate these issues (Bickley and Torgler 2023).

### 5.2.1. Twinning mini-publics

To illustrate how this approach could work in practice, consider developing a digital twin (DT) of a mini-public—a small group of city residents convened to deliberate on a policy issue, such as how to allocate funding for environmental initiatives or how to address local traffic challenges. Real mini-publics of this nature exist in practice; for example, the UK Climate Assembly in 2020 (Willis, Curato, and Smith 2022) is a well-known case. These deliberative forums are often chosen for experimental methodologies (Grönlund and Herne 2022) due to their manageable size and the potential for structured evaluation. Typically, the main objective is to produce policy recommendations.

[4] It is not strictly necessary that "all" what-if simulations must use ML to predict changes—one could also use rule-based approaches or simpler parameter sweeps.

Within a mini-public, constructive dialogue can be measured by the depth of argumentation (including evidence or references) and balance in speaking time (Steenbergen et al. 2003; André Bächtiger, Gerber, and Fournier-Tombs 2022; Scudder 2022). Accordingly, agent behavior in the DT simulation might reflect the likelihood of participants bringing new evidence, responding to opposing views, or remaining silent. Meanwhile, SNA indicators can gauge how ideas spread and identify dominant groups or influencers, assessing whether this dominance affects argumentation depth or pluralism positively or negatively (Musso and Helbing 2024). The constructiveness of a debate may be assessed through sentiment analysis.

Ensuring the demographic representativeness of the mini-public is often complex (Germann 2025). Still, DT designers can gather institutional data (for population attributes) and engagement platform data (for behaviors and interactions) to perform what-if testing of different compositions to see if they yield outcomes like those you would expect from the broader population. When building an ABM or other simulation techniques, each agent is randomly assigned demographic attributes that align with official city-wide distributions (if direct participant-level data are unavailable). Engagement-platform metrics (e.g., number of posts, comments, votes, timestamps, and discussion content) help distinguish between (the spectrum of) more active and passive agents. Sentiment analysis or topic modeling can reveal whether participants favor or oppose particular policies.

Based on these data, agent profiles can be constructed to represent participants' behaviors and general orientation. Crucially, the simulation does not need to replicate a specific community exactly, as the goal is not twinning or replacing human constituents. It can be a hypothetical or randomized (i.e., synthetic) group whose characteristics can derive or match distributions, correlation and interactions of real-world populations faithfully while preserving privacy. Also, this facilitates sensitivity studies about potential risks and biases caused by the selection process of participants in mini-publics in real life. This approach follows data protection regulations and ethical standards (Bäumer et al. 2024), according to which personal data must be anonymized or pseudonymized, so individuals cannot be re-identified, while retaining the demographic and behavioral variations necessary for a meaningful simulation.

Once the data have been cleaned, they are integrated into the mini-public's DT. The core model may use ABM, which involves assigning each agent (representing a participant) specific attributes such as age, income, and education. Decision rules that govern agent behavior must be defined to mirror real-world participant dynamics, such as the probability of posting versus lurking or the likelihood of shifting positions when presented with well-argued evidence.

Each of these artificial agents is a human DT of a synthetic population that is statistically equivalent to the real population. Depending on the specific case, different anonymization techniques may be used.

The ABM also requires the design of interaction rules based on behaviors derived from digital platforms, such as posting and commenting patterns. For example, the model may simulate how agents communicate in discussion threads, form temporary coalitions, respond to arguments, or change their opinions. For instance, if an agent with moderate environmental concerns encounters a well-supported argument from a highly respected participant or facilitator, the level of concern may slightly increase. Similarly, if real-world data show that participants from certain demographic groups tend to cluster or share similar views, the ABM can replicate this behavior by increasing the "likelihood to align with similar agents".

The same cleaned data—particularly those about participation trends—can also inform a System Dynamics (SD) model. This is not trivial, as in SD, individual-level data become stocks and flows that capture system-wide dynamics. This approach is useful for modeling higher-level feedback loops, such as how increased dissatisfaction can reduce future participation or how extended debate might simultaneously promote consensus and lead to participant fatigue.[5]

Real-world data on interactions—such as mentions, replies, or co-annotations—can be used to construct adjacency matrices for Social Network Analysis (SNA). These matrices represent the structure of a network, where nodes represent individuals and edges represent interactions. Once the network structure is established, it can be analyzed to measure centrality (identifying key influencers), detect subgroups (e.g., communities or clusters), and track idea propagation (e.g., how information spreads). SNA can help identify potential influencers within a group, assuming their behavior in the simulation reasonably mirrors their real-world actions, such as in a social media context. It can also reveal how ideas propagate between subgroups, highlighting phenomena like bridging connections (where subgroups interact) or echo chambers (where subgroups become isolated).

Finally, DES can be employed to model how specific events alter the state of the mini-public. For example, the timing of real-world actions—such as when participants post, vote, or change their stances—can be used to define events and event triggers in the simulation. Observed frequencies from actual data (e.g., "a major influencer emerges in 1 out of 3 debates") can be translated into probabilities within the DES. This allows the simulation to replicate realistic patterns of behavior and interaction. Predictions about future community states—such as who is likely to drop out, who will submit or rebut arguments, or how likely a proposal will succeed—often require Machine Learning, Natural Language Processing, and sentiment analysis. Sentiment analysis, in particular, can quantify changes in support, neutrality, or opposition (Liu 2022).[6]

[5] In practice, going from raw, cleaned data to a fully calibrated ABM or SD model often requires more intermediate steps: e.g., sensitivity analysis to test how changes in parameters affect outcomes (Ligmann-Zielinska et al. 2014).

In short, following state-of-the-art societal and behavioral simulation, one of the most granular possibilities for generating a DT for a mini-public can result in a multi-stage pipeline integrating structured socio-demographic data, behavioral modeling, and LLM-based personas for the AI agents. These autonomous agents can populate a dynamic environment, in which they will interact and exchange information dynamically with other agents and with a changing context (e.g., policies, protocols, news), as shown in Figure 1. These changing conditions will also inform their personas by updating their memory and opinion based on actions and answers. Aggregated metrics at the emerging network can be extracted and used to calibrate and evaluate the performance of the digital twin over historical data.

[6] Because this is still a simulation (even in "digital twin" form), it must be calibrated and validated to ensure the agents' collective behavior closely matches real-world patterns. One way to achieve this is to simulate a known past event (e.g., last year's participatory budgeting process) and compare the predicted outcomes (such as final proposal support rates) to the actual historical results. This calibration step helps to improve the DT's predictive accuracy and is a precondition for its validation.

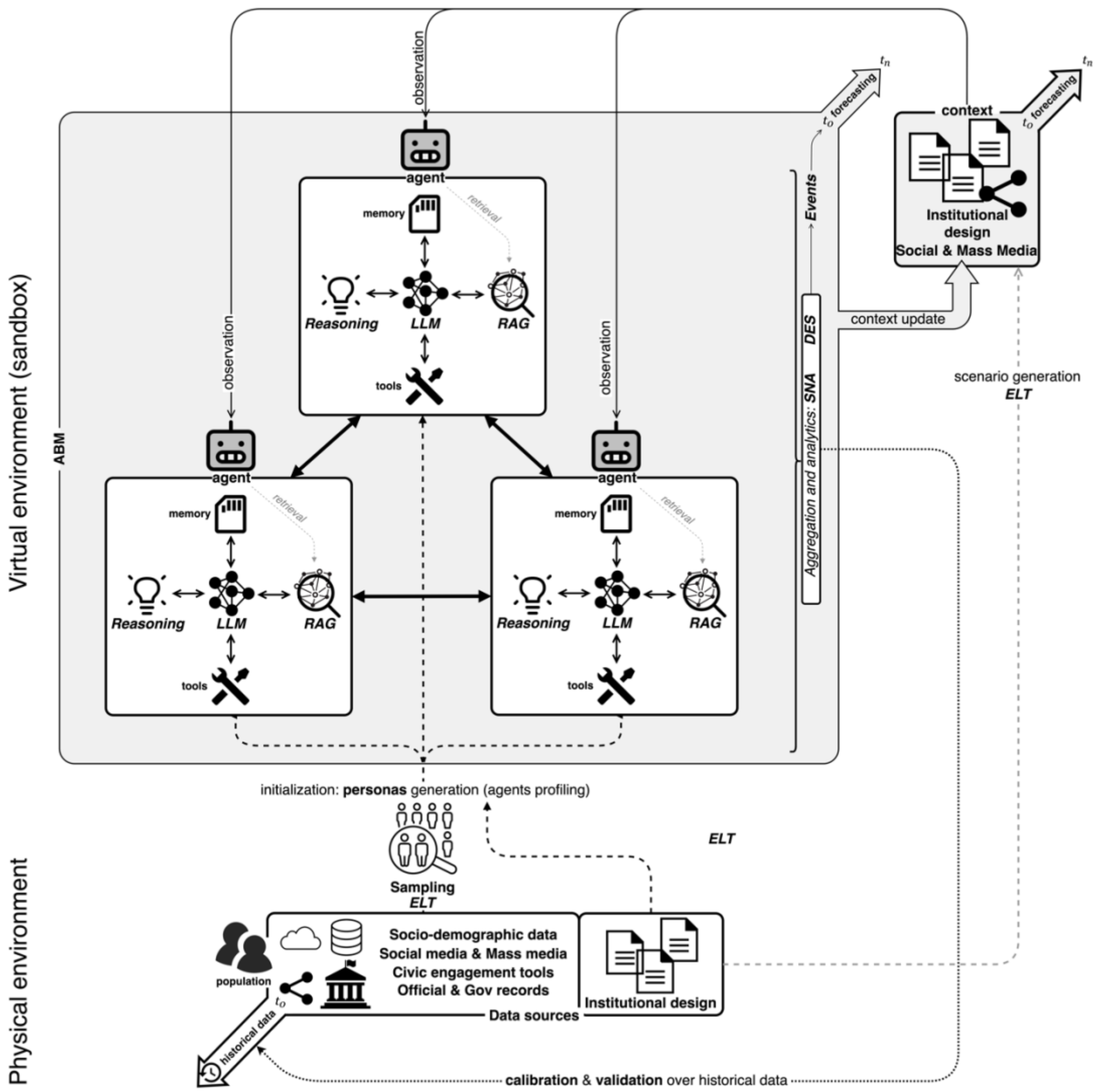

*Figure 1. Proposed Multi-agent-based model (ABM) architecture for a democracy deliberation DT, based on AI-powered agents profiled as different personas based on privacy-preserving synthetic populations that statistically match actual socio-demographic, behavioral, and public records data, which is pre-processed through extract, load, and transform (ELT) processes. Each agent is equipped with multiple modules able to perform different tasks (reasoning, information retrieval, memorizing, action planning) that allow them to communicate and interact with other agents and the context (e.g., institutional design and news), and therefore, obtaining feedback to adjust to the changing environment. The virtual environment exhibits emerging social network features that can then be related to social network analytics (SNA). The aggregation of these social dynamics can be used for higher-scale modeling, such as Discreet-Event Simulation (DES), to calibrate and evaluate the performance of the DT based on actual historical data (as layers of data sources) and to update the context over time, forecasting new scenarios.*

### 5.3. Using the DT to test (and refine) deliberative democracy rules

Our objective is not to replace real deliberative communities by digital replicas. Substituting human agents with virtual approximations would lead to a scenario fraught with oversimplifications. In particular, it would not represent the diversity of opinions well. Instead, as noted earlier, the goal is to use the DT as a sandbox for experimenting with procedural rules. By running what-if scenarios, we can examine how institutional designs unfold and assess whether they achieve their stated outcomes.

DT may help deliberative democracy designers to conduct experiments that could not be performed in the real world (Helbing and Sánchez-Vaquerizo 2023, 65), especially given the constraints discussed in Section 3.1. To illustrate this, we turn to the three main sets of procedural rules outlined in Section 3.

a) *Pre-deliberation rules.* Pre-deliberation rules address the composition of participants and other preparatory conditions for deliberation. A foundational aspect of these rules is determining who should participate. Here, a DT can be employed to evaluate different recruitment strategies and their impact on inclusivity, diversity, and the overall quality of deliberation. Through ABM—simulating virtual participants with varied demographics, interests, and needs—designers can explore how different selection mechanisms (e.g., random vs. weighted sampling) influence group composition and discussion dynamics. For example, simulations can reveal whether purely random selection yields a diverse group representation or whether specific demographics are likelier to drop out due to low interest or external constraints, thereby reducing actual diversity. If the goal is to represent real diversity or even to maximize it, results might indicate that simple random selection is inadequate because of high opt-out rates or systematic underrepresentation of minorities, thus favoring weighted or stratified sampling (Rowe and Frewer 2000; Smith 2009). DT can precisely quantify the impact of this under various conditions. Combining representative sampling with minority-focused panels can also address challenges related to small sample sizes. Conversely, if the goal is to increase public buy-in, simulations can be used to evaluate voluntary participation as a recruitment strategy. They might show that volunteers are more motivated to become informed and engage in reasoned discussion; however, they could also reveal that this approach disproportionately attracts individuals with strong pre-existing opinions, indicating a need for neutral outreach campaigns to broaden participation.

   Through the same simulation approaches, we can compare stakeholder-targeted recruitment (e.g., inviting specific groups such as activists or business leaders) with more open or random selection to see whether it promotes balanced participation or leads to dominance of specific interests or increased polarization. DES can further reveal how financial incentives or mandatory participation policies influence participation rates.

Beyond composition, the quality of deliberation can serve as a proxy for evaluating how well different groups are represented. For instance, measuring “argument diversity”—using metrics such as entropy indices or the number of distinct argument types (Palau and Moens 2009; Grimmer and Stewart 2013)—can help determine whether including minority voices meaningfully improves debates or if they remain overshadowed in practice.

Another key set of pre-deliberation rules involves informational preparation on the topics to be discussed. Agent-Based Modeling (ABM) and Social Network Analysis (SNA) can be used to understand how participants absorb information and how it spreads within subgroups. For instance, SNA could construct interaction graphs to track information flow and identify influential participants. The objective here should not be to target or suppress influential voices—an approach often associated with totalitarian regimes or non-pluralistic platforms—but rather to reduce influence concentration. Promoting a more balanced influence distribution can help prevent uncritical epistemic deference and encourage more independent, reasoned participation.

Likewise, DES can help determine how long the preparatory phase should be: e.g., simulations might show that expert presentations increase overall knowledge but risk to create excessive deference to authority (although deference may be tricky to quantify in the simulation), whereas encouraging participants to do background readings may foster deeper, self-driven learning. The optimal approach will depend on the primary goal—whether it is factual accuracy, broad participation, or some balance of the two—and might include evidence-oriented measures to mitigate the risk of undue deference.

Representativeness is often an ideal rather than a fully attainable standard. Designers must balance the desire for diverse participation with real-world constraints like budget, recruitment time, and participant availability. Here, a DT can simulate how these constraints shape different selection rules. DES or system dynamics can project how much recruitment time or funding each rule requires, helping designers refine their choices to maximize representativeness without exceeding practical limits.

Finally, agendas could also be mapped using computational methods. Sentiment analysis, for example, can gauge how different agendas might influence participant engagement or emotional reactions by analyzing data from simulated interactions, while ABM can reveal whether a collaboratively developed agenda generates greater ownership despite taking longer to finalize.

b) *Discussion rules*. The design of discussion formats can be enhanced with DT simulations (e.g., digital twins or comparable digital tools) to test various session structures. For example, simulations can help determine whether deeper argumentation is best achieved in a single session or multiple rounds (e.g., iterating homogeneous and heterogeneous panels/sessions), examining how argument complexity evolves. Argument complexity can be measured using metrics such as layers of reasoning (e.g., the number of premises and rebuttals in an argument) or the degree of justification (e.g., the extent to which arguments are backed by evidence and acknowledged counterpoints) (Walton, Reed, and Macagno 2008; Prakken and Vreeswijk 2002; Lippi and Torroni 2016).

Simulations might reveal that single-session formats work better if the goal is efficient deliberation. However, if the goal is to reduce polarization or bias, simulations may suggest that iterative deliberation—with repeated exposure to opposing views (Yang and Bachmann 2025)—is more effective. That said, recent ABM for deliberative democracy (Lee et al. 2022) shows that, while multiple rounds of discussion are more impactful than simply providing information, their incremental utility decreases over time, with limited effects on reducing polarization.

Simulations can also test how participants present opinions and rebut arguments and how these modalities affect deliberation outcomes such as fairness or engagement. For example, DES can model how different time constraints impact argument complexity or completeness. Additionally, ABM combined with NLP can analyze whether rebuttals deepen discussions or lead to repetition. System Dynamics can further simulate how rebuttal styles (e.g., iterative step-by-step refinements vs. one-time paired rebuttals) influence opinion shifts and consensus-building.

Moreover, since moderators or facilitators strongly affect discussion dynamics, DT simulations can evaluate which facilitation styles best promote argument diversity and fairness and whether frequent moderator interventions enhance or hinder the natural flow of discourse. SNA can help detect power imbalances, such as a facilitator who repeatedly calls on the same participants.

A critical aspect of discussion rules concerns how final decisions are reached. Different decision-making strategies can be tested via ABM, possibly supported by game-theoretic models that treat participants as rational or boundedly rational agents, employing strategic behavior rather than simply sharing sincere opinions. These models can show how power imbalances and knowledge asymmetries shape coalition success (Parsons, Gymtrasiewicz, and Wooldridge 2012). More specifically, we may simulate whether participants adopt cooperative or aggressive strategies when debating contentious issues (the so-called hawk-dove game) and evaluate whether conflict resolution rules (structured mediation, facilitation styles) encourage cooperation (Amadae and Watts 2023).

Through these simulation techniques, designers can draw insights about the trade-offs of different decision-making rules. So, for instance, DT simulations can reveal that consensus-building, while fostering broader participation and depth of dialogue, can also slow down decision-making. Conversely, majority voting may speed up decision processes, but risks systematically disadvantaging certain groups—perhaps due to unequal cooperation or strategic abilities—resulting in disengagement and reduced overall effectiveness of deliberation (Yang, Hausladen, et al. 2024).

c) *Post-deliberation rules*. The deliberative process also includes rules for monitoring and implementing decisions after deliberation, though these are often less emphasized than pre-deliberation or in-process rules. Unlike the other two sets of procedural rules, which can be tested using simulated data from online interactions, post-deliberation rules mostly require real-world data sources. This focus on real-world data is necessary because post-deliberation interactions are highly context-specific—lacking reliable proxies in other datasets—and difficult to replicate in simulations. Accordingly, the emphasis shifts to data analytics more than simulation. These data include surveys, voting records, and historical records of participatory policymaking. Such data help researchers understand who engages in deliberation and how participants react to post-deliberation processes.

In this context, data analytics tools can be used to determine the most feasible documentation and reporting mechanisms. For example, these tools can help evaluate whether open-access reports increase public trust or lead to selective misinterpretation compared to restricted reports. Additionally, they can analyze historical feedback data to assess how feedback collection methods—such as anonymous vs. named feedback, immediate vs. delayed feedback, or facilitated group reflections—impact participant engagement and depth of reflection.

Sentiment Analysis and Natural Language Processing can provide insights into participant engagement, though qualitative analysis should supplement their interpretations. These tools can reveal how participants feel about the deliberation process and what they focus on when giving feedback or discussing outcomes. So, for instance, if people who saw the open-access reports consistently express more positive sentiment (e.g., relief, trust, willingness to engage) compared to those who saw the restricted reports, it suggests open access may foster greater public confidence. Conversely, if negative sentiments (e.g., anger, skepticism) dominate when reports are fully open, it might indicate issues like data overload or misunderstandings.

Before we move on to the potential advantages and limitations of this "sandboxing" strategy, it is crucial to note that, although we used an illustrative example of a single DT replicating an entire community and its institutional design, in practice, this may not be the most feasible approach. A DT may also be used to test only specific aspects that designers wish to prioritize. For instance, if the recruitment strategy is already well-defined, designers might use the DT solely to examine how best to introduce new arguments during discussions or how final decisions should be made, rather than modeling every phase of the process.

## 6. Limitations and future research

Digital twins offer a promising avenue for empirical research into procedural rules, particularly given the challenges of replicating, iterating, and scaling deliberative experiments in physical or lab settings. However, their effectiveness depends on the accuracy of behavioral assumptions, the quality of input data, and their capacity to generalize to real-world democratic practice. For instance, oversimplified assumptions about specific agent behaviors and decision-making rules (e.g., how and when participants disengage) can introduce biases and lead to misleading conclusions. Moreover, simulating future outcomes may be challenging because of the inherent complexity involved in human-based and social systems. For example, the complex dynamics observed in networked systems are challenging to model, fit, and validate (Caldarelli et al. 2023).

These limitations highlight trade-offs and challenges of creating a digital twin of a city or society, including its people, to explore and assess different democratic deliberation and voting formats. A complete and accurate digital replica of every aspect of society may be infeasible, calling for a scale-down and developing issue-specific digital twins.

We realize, however, that digital twins have been in the making already for some time, not always with the purpose of modeling and assisting democracies with digital technologies. In principle, digital technologies can also be used to support other data-driven, AI-managed, cybernetic forms of society, ranging from digital versions of feudalism, communism, or fascism to other forms of digital societies.

Moreover, as observed for DTs of cities (Batty 2021), DTs exist on a range between strictly mirroring the physical environment (which limits speculative exploration) and decoupled simulations (that enable testing of alternative scenarios). This means that, for democratic deliberation, DTs may be used as predictive models (running before any phase in the deliberation) or as real-time interactive systems (that assist and evolve alongside the physical, human-driven processes). This also enables such frameworks to support different types of agencies, processes, and involvement levels from humans, including hybrid, collective, and symbiotic intelligence of humans and AI (e.g., Human-in-the-Loop).

Key steps for future research to ensure that the use of digital twins aligns with the values of (liberal and deliberative) democracies are:

- *Avoid surveillance*. Effective measures should be taken to prevent personal surveillance, scoring, and targeting. This means that one should not strive to create and use as-identical-as-possible digital twins of people, but "noisy" digital twins, i.e., to work with samples of hypothetical personas having statistically representative (or theoretically assumed) characteristics. Hence, we propose to work with privacy-preserving synthetic data—a statistical digital twin ("[approximate] replica") with the same distribution of data of relevant variables (including comparable correlations). A favorable side effect of this is that the outcomes are expected to be less dependent on random factors and events and more resilient to disruptions.

- *Ensure transparency and accountability*. Ensuring transparency in the data, software, and procedures is crucial to getting people's trust and public support and encouraging participation. Otherwise, there could be hidden manipulation. Transparency is also essential for ensuring accountability in all aspects of digital deliberation—whether in the design and implementation of deliberative practices or in addressing concerns and making necessary corrections.

- *Open, plural, and fair discussion*. The deliberation platform should offer a level playing field such that the competition of ideas and the consideration of justified individual interests are fair. It should promote the mutual understanding of different perspectives. In doing so, it should support a respectful exchange of ideas and a constructive dialogue, which allows people to voice diverse opinions, interests, and needs without fear.

- *Avoid mis- and disinformation*. The platform should promote accurate, verifiable information rather than mis- or disinformation. It should support an evidence-based approach emphasizing critical evaluation, hypothesis verification, and falsification rather than relying solely on raw data, which can be misinterpreted or manipulated without proper contextual analysis.

- *Human-centered approach*. The platform should have a human-centered approach that serves the interests and needs of people (rather than merely optimizing efficiency, automation, or institutional control). Given that AI often functions as an opaque “black box” and inherently complex social systems, a human-centered approach should integrate cognitive architectures that mimic human thought processes (instead of relying solely on data-driven methods).

- *Richness and multi-dimensionality of values*. The platform should consider legal, ethical, societal, and cultural values and qualities people care about, such as human dignity, friendship, love, trust, creativity, beauty, etc.

- *Enhance human agency*. The platform should support human agency and offer participatory opportunities that allow people to co-create solution ideas.

In short, digital twin technology should serve as an open and inclusive platform rather than be used for “control room” purposes that aim at socially engineering people’s settings, environments, and behaviors. In this way, multiple stakeholders could openly exchange perspectives, jointly address conflicts, and simulate solutions in a virtual space before implementing concrete actions in the physical world, aligned with the 'peace room' approach (Helbing and Seele 2017).

Last but not least, DT technology can ultimately help find alternatives to the traditional models of deliberative democracy that we have discussed in this paper, i.e., based on structured debates and formal decision-making processes. These conventional methods often follow linear, rule-based, and argumentative structures. In contrast, DTs and AI may enable more dynamic, flexible approaches, such as re-mixing. In fact, instead of locking participants into strict plans and fixed positions, remixing is an iterative decision-making process that allows people to submit modular elements of their views and combine and modify them also based on real-time feedback. In short, ideas would (co-)evolve based on feedback and experimentation, avoiding premature commitment to a single plan. DTs offer an environment where re-mixing can be simulated, adjusting the decision collaboratively before real-world implementation.